\documentclass[11pt,twoside,a4paper,cmspaper,final,collab]{cms-tdr}
\def\svnVersion{705735e-D}\def\svnDate{2025/05/15}\def\cmsCernNoTag{CERN-EP-2024-336}\def\cmsCernDate{\today}\def\cmsMessage{Published in the Journal of High Energy Physics as \href{http://dx.doi.org/10.1007/JHEP05(2025)079}{\doi{10.1007/JHEP05(2025)079}.}}
\begin{document}\cmsNoteHeader{HIG-24-013}

\newcommand{\fb}{\unit{fb}}

\newcommand{\usedLumiRunIII}{34.7\fbinv\xspace}
\renewcommand{\pt}{\ensuremath{p_{\text{T}}}\xspace}
\newcommand{\mll}{\ensuremath{m_{\Pell\Pell}}\xspace}
\newcommand{\ggH}{\ensuremath{\Pg\Pg\to\PH}\xspace}
\newcommand{\ttH}{\ensuremath{\cPqt\PAQt\PH}\xspace}
\newcommand{\qqZZ}{\ensuremath{\PQq\PAQq\to\PZ\PZ}\xspace}
\newcommand{\ggZZ}{\ensuremath{\Pg\Pg\to\PZ\PZ}\xspace}
\newcommand{\HZZfl}{\ensuremath{\PH\to\PZ\PZ\to4\Pell}\xspace}
\newcommand{\Hllll}{\ensuremath{\PH\to4\Pell}\xspace}
\newcommand{\mH}{\ensuremath{m_{\PH}}\xspace}
\newcommand{\mllll}{\ensuremath{m_{4\Pell}}\xspace}
\newcommand{\fmu}{\ensuremath{4\PGm}\xspace}
\newcommand{\fe}{\ensuremath{4\Pe}\xspace}
\newcommand{\tetmu}{\ensuremath{2\Pe2\PGm}\xspace}

\cmsNoteHeader{HIG-24-013}
\title{Measurements of the Higgs boson production cross section in the four-lepton final state in proton-proton collisions at \texorpdfstring{$\sqrt{s} = 13.6\TeV$}{sqrt(s) = 13.6 TeV}}
\date{\today}

\abstract{
The measurements of the Higgs boson (\PH) production cross sections performed by the CMS Collaboration in the four-lepton ($4\Pell, \Pell=\Pe,\PGm$) final state at a center-of-mass energy $\sqrt{s}=13.6\TeV$ are presented. These measurements are based on data collected with the CMS detector at the CERN LHC in 2022, corresponding to an integrated luminosity of 34.7\fbinv. Cross sections are measured in a fiducial region closely matching the experimental acceptance, both inclusively and differentially, as a function of the transverse momentum and the absolute value of the rapidity of the four-lepton system. The $\PH\to\PZ\PZ\to 4\Pell$ inclusive fiducial cross section is measured to be $2.89^{+0.53}_{-0.49}\stat^{+0.29}_{-0.21}\syst\unit{fb}$, in agreement with the standard model expectation of $3.09^{+0.27}_{-0.24}\unit{fb}$.}

\hypersetup{%
pdfauthor={CMS Collaboration},%
pdftitle={Measurements of Higgs boson production cross section in the four-lepton final state in proton-proton collisions at sqrt(s)=13.6 TeV},%
pdfsubject={CMS},%
pdfkeywords={CMS, Higgs boson, Run 3, cross section}} 

\maketitle

\section{Introduction}
\label{sec:intro}
The discovery of the Higgs boson (\PH) in 2012 by the ATLAS and CMS Collaborations~\cite{Aad:2012tfa,Chatrchyan:2012ufa,Chatrchyan:2013lba} was a major milestone and provided a crucial missing element of the standard model (SM) of particle physics.
Since then, many measurements of the Higgs boson properties~\cite{ATLASPropertiesRun1,CMSPropertiesRun1,ATLASCMSMassRun1,ATLASCMSPropertiesRun1,ATLASPropertiesRun2,CMSPropertiesRun2} have been performed which further validate the SM predictions.

The four-lepton decay channel (\HZZfl, $\Pell=\Pe,\PGm$) played a very important role in the discovery of the Higgs boson in 2012. Despite its low branching fraction ($1.3\times 10^{-4}$), it benefits from a clear four lepton signature, large signal-to-background ratio, and the possibility to fully reconstruct the final-state kinematics.
Thanks to these characteristics, the four-lepton final state has been extensively used to measure the Higgs boson properties.
The measurements performed include the determination of the Higgs boson mass (\mH), spin and parity~\cite{ATLASH4lLegacyRun1,CMSH4lLegacyRun1,CMSH4lSpinParity,CMSH4lAnomalousCouplings,CMSH4l2016,ATLASH4l2016, CMSHIG21019},
width~\cite{CMSH4lWidth,CMSH4lLifetime,ATLASH4lWidth,ATLASH4lWidth2016}, inclusive and differential fiducial cross sections~\cite{ATLASH4lFiducial8TeV,CMSH4lFiducial8TeV, CMSH4l2016,ATLASH4lFiducial2016, ATLASH4lLegacyRun2, ATLASH4lFiducialRun2, CMSHIG21009}, as well as the tensor structure for interactions with a pair of gauge bosons~\cite{CMSH4lAnomalousCouplings,CMSH4lLifetime, CMSH4lAnomalousCouplings2016,ATLASH4l2016,CMSHVVAnomalousCouplings2016,CMSHIG19009}.

Fiducial cross section measurements are one of the most common approaches for the characterization of the Higgs boson production and decay.
They allow less model-dependent results than total cross sections.
The ATLAS and CMS Collaborations measured fiducial cross sections in the $\PH\to\PGg\PGg$~\cite{ATLAS:2022fnp,CMS:2022hyj}, $\PH\to\PW\PW$~\cite{ATLAS:2023hyd,ATLAS:2023pwa,CMS:2020dvg}, \HZZfl~\cite{ATLASH4lFiducialRun2,CMSHIG19001,CMSHIG21009}, and $\PH\to\text{bb}$~\cite{ATLAS:Hbb,CMS:Hbb} decay channels using the full data set collected in 2016, 2017, and 2018.
The CMS Collaboration also reported results in the $\PH\to\PGt\PGt$~\cite{CMS:2021gxc} decay channel and their combination with the $\PH\to\PGg\PGg$, $\PH\to\PW\PW$, and \HZZfl decay channels~\cite{CMS-PAS-HIG-23-013},
while the ATLAS Collaboration presented results from the combination of the $\PH\to\PGg\PGg$ and \HZZfl decay channels, both with data at $13\TeV$~\cite{ATLAS:2022qef} and 13.6\TeV~\cite{ATLAS:run3}.

The measurement of the Higgs boson fiducial cross section as a function of the center-of-mass energy is an important test of the SM.
This paper reports the inclusive and differential fiducial cross sections for Higgs boson production in the \HZZfl decay channel using data from proton-proton ($\Pp\Pp$) collisions at $\sqrt{s}=13.6\TeV$ recorded with the CMS detector at the CERN LHC in 2022 and corresponding to an integrated luminosity of \usedLumiRunIII~\cite{CMS-PAS-LUM-22-001}.
The fiducial phase space region is defined to closely reproduce the experimental acceptance and reconstruction-level selection criteria, to reduce the model dependence of the results.
The differential cross sections are measured as a function of the transverse momentum ($\pt^{\PH}$) and absolute value of the rapidity ($\abs{y_{\PH}}$) of the Higgs boson.
The analysis relies on the methods that have been optimized in previous studies to characterize the Higgs boson properties in the four-lepton decay channel~\cite{CMSH4lFiducial8TeV, CMSHIG19001,CMSHIG21009}.

This paper is organized as follows.
The CMS detector is briefly described in Section~\ref{sec:detector}.
The data and simulation used are discussed in Section~\ref{sec:samples}.
The event reconstruction techniques and selection criteria are detailed in Section~\ref{sec:selection}.
The background estimation is discussed in Section~\ref{sec:bkgd} and the definition of the fiducial phase space region where the differential cross sections are measured is presented in Section~\ref{sec:fiducialvolume}.
The signal modeling and statistical procedure adopted in the extraction of the results are described in Section~\ref{sec:measurement}.
The systematic uncertainties that affect the measurement are described in Section~\ref{sec:systematics}.
The results are presented in Section~\ref{sec:results}, followed by a summary in Section~\ref{sec:summary}.

\section{The CMS detector}
\label{sec:detector}
The central element of the CMS experiment is a superconducting solenoid providing a magnetic field of 3.8\unit{T}. A silicon pixel and strip tracker, a lead tungstate crystal electromagnetic calorimeter (ECAL), and a brass and scintillator hadron calorimeter (HCAL) are placed within the solenoid volume and each of them is composed of a barrel and two endcap sections.
Forward calorimeters extend the pseudorapidity ($\eta$) coverage provided by the barrel and endcap sections. Muons are reconstructed using gas-ionization detectors embedded in the steel flux-return yoke outside the solenoid.
The CMS experiment relies on a two-tiered trigger system to select the events of interest.
The first level is composed of custom hardware processors and selects events
at a rate of approximately 100\unit{kHz} within a fixed latency of about 4\mus~\cite{Sirunyan:2020zal} using information from the calorimeters and muon detectors.
The second level, called high-level trigger, consists of a farm of processors running a version of the full event reconstruction software optimized for fast processing, and reduces the event rate to around 5\unit{kHz} before data storage~\cite{CMS:2024aqx}.
A more detailed description of the CMS detector, together with the definition of the coordinate system used and the relevant kinematic variables, can be found in Refs.~\cite{Chatrchyan:2008zzk,CMS:2023gfb}.

\section{Data and simulated samples}
\label{sec:samples}
Signal samples are simulated at next-to-leading order (NLO) in perturbative quantum chromodynamics (pQCD) using the \POWHEG~2.0~\cite{Nason:2004rx,Frixione:2007vw,Alioli:2010xd} generator for the main production mechanisms of the SM Higgs boson: gluon fusion ($\ggH$)~\cite{Alioli:2008tz} including quark mass effects~\cite{Bagnaschi:2011tu}, vector boson fusion (VBF)~\cite{Nason:2009ai}, associated production with a vector boson (VH, where \PV = \PW, \PZ)~\cite{Luisoni2013}, and associated production with a pair of top quarks ($\ttH$)~\cite{Hartanto:2015uka}.
Events produced via the $\ggH$ mechanism are reweighted to match the predictions at next-to-NLO order (NNLO) in pQCD, including parton shower matching (NNLOPS)~\cite{Hamilton:2013fea} as functions of $\pt^{\PH}$ and the jet multiplicity in the event.
The Higgs boson production in association with $\PQb$ quarks is not considered as its contribution is negligible with respect to the other production modes.
The decay of the Higgs boson to four leptons is modeled with JHUGEN~7.0.2~\cite{Gao:2010qx, Bolognesi:2012mm,Anderson:2013afp,Gritsan:2016hjl,Gritsan:2020pib}. All samples are generated with the NNPDF 3.1 NNLO parton distribution
functions (PDFs)~\cite{Ball:2014uwa}.
The simulation of the various production and decay modes is based on the theoretical predictions from Refs.~\cite{Anastasiou:2015ema,Anastasiou2016,Ciccolini:2007jr,Ciccolini:2007ec,Bolzoni:2010xr,Bolzoni:2011cu,Brein:2003wg,Ciccolini:2003jy,Beenakker:2001rj,Beenakker:2002nc,Dawson:2002tg,Dawson:2003zu,Yu:2014cka,Frixione:2014qaa,Demartin:2015uha,Demartin:2016axk,Denner:2011mq,Djouadi:2018xqq,Bredenstein:2006rh,Bredenstein:2006ha,Boselli:2015aha,Actis:2008ts}, which are summarized in Ref.~\cite{deFlorian:2016spz}.
The signal samples are normalized to the cross sections provided by the LHC Higgs Working Group for $\sqrt{s}=13.6\TeV$~\cite{Karlberg:2024zxx}.

The main background originates from $\PZ\PZ$ production via quark-antiquark annihilation and gluon fusion. The former is simulated at NLO in pQCD with \POWHEG~2.0~\cite{Melia:2011tj}, while the latter is generated at leading order (LO) with \MCFM~7.0.1~\cite{MCFM,Campbell:2011bn,Campbell:2013una,Campbell:2015vwa}.
The production via quark-antiquark annihilation is reweighted to NNLO using a $K$ factor computed as a function of $m_{\PZ\PZ}$, relying on the NNLO computation of the $\qqZZ$ fully differential cross section~\cite{Grazzini2015407}.
An additional NLO electroweak correction is applied as a function of $m_{\PZ\PZ}$ according to the computation presented in Ref.~\cite{Bierweiler:2013dja}.
Although no exact calculation exists beyond LO for the \ggZZ background, it has been shown~\cite{Bonvini:1304.3053} that the soft-collinear approximation is good enough to describe the background cross section and the interference term at NNLO\@. Further calculations also show that the $K$ factors are very similar at NLO for the signal and background~\cite{Melnikov:2015laa} and at NNLO for the signal and interference terms~\cite{Li:2015jva}.
This suggests that the same $K$ factor can be used for the $\Pg\Pg\to\PH\to\PZ\PZ$ signal and \ggZZ background~\cite{Passarino:1312.2397v1}.
The current analysis relies on an NNLO/LO $K$ factor computed for the signal process as a function of the di-boson invariant mass $m_{\PZ\PZ}$, using the \textsc{hnnlo} v2 program~\cite{Catani:2007vq,Grazzini:2008tf,Grazzini:2013mca}.
The NNLO and LO $\Pg\Pg\to\PH\to 2\Pell 2\Pell'$ cross sections are computed for a fixed Higgs boson decay width of 4.07\MeV across the whole $m_{\PZ\PZ}$ mass range, and the $K$ factor is defined as their ratio.
The $K$ factor is then applied to the \ggZZ background process as a function of $m_{\PZ\PZ}$. We note that the $K$ factors used in this analysis are identical to those of the Run~2 analysis at $\sqrt{s}=13\TeV$~\cite{CMSH4l2016} since no noticeable difference is expected between the two energies in the $m_{\PZ\PZ}$ mass range of relevance.

The additional background contribution arising from the production of \PZ bosons with associated jets ($\PZ$+jets) is estimated with the technique based on control samples in data, as described in Ref.~\cite{CMSHIG19001} and in Section~\ref{sec:redbkgd}.
The $\PZ$+jets process is simulated at NLO with the \MGvATNLO 2.4.2~\cite{amcatnlo} program and is used for validation studies, for the training of the boosted decision tree (BDT) adopted for the identification and isolation requirements on electrons, and for the derivation of data to simulation scale factors.

The Monte Carlo (MC) generators are interfaced with \PYTHIA version 8.230~\cite{Sjostrand:2014zea} using the CP5 underlying event tune~\cite{Sirunyan:2019dfx} to simulate the parton showering and hadronization effects.

All MC generated events are processed through a simulation of the CMS detector based on \GEANTfour~\cite{Agostinelli:2002hh,GEANT} and are reconstructed with the same algorithms used for the data. Additional pp interactions in the same and nearby bunch crossings, referred to as pileup, are also simulated. The distribution of the number of pileup interactions in the simulation is adjusted to match that observed in data.

\section{Event reconstruction and selection}
\label{sec:selection}
Candidate events are required to have charged leptons passing loose identification and isolation requirements~\cite{CMS:2020uim,Sirunyan:2018}, following the online selection based on dielectron, dimuon, and electron-muon high-level trigger algorithms.
Triggers that require either a single lepton or three leptons are also used to increase the efficiency, which is larger than 99\% for the events that satisfy the selection requirements presented in what follows, and is in agreement with that estimated from simulated samples.
The trigger efficiency is determined following the same procedure as in Ref.~\cite{CMSHIG19001}.

To reconstruct and identify particles in an event, the particle-flow (PF) algorithm~\cite{CMS-PRF-14-001} is used, which employs an optimized combination of information from the various elements of the CMS detector.
The energy of photons is obtained from the ECAL measurement. The energy of electrons is determined from a combination of the electron momentum at the primary vertex (PV), as determined by the tracker, the energy of the corresponding ECAL cluster, and the energy sum of all bremsstrahlung photons spatially compatible with originating from the electron track.
The PV is taken to be the vertex corrsponding to the hardest scattering in the event, evaluated using tracking information alone, as described in Section 9.4.1 of Ref.~\cite{CMS-TDR-15-02}.
The momentum of muons is obtained from the combined information of the tracker and the muon chambers. The energy of charged hadrons is determined from a combination of their momentum measured in the tracker and the matching ECAL and HCAL energy deposits, corrected for the response of the calorimeters to hadronic showers. Finally, the energy of neutral hadrons is obtained from the corresponding corrected ECAL and HCAL energy deposits.

Muons with $\pt^{\PGm} > 5\GeV$ and $\abs{\eta^{\PGm}} < 2.4$ are reconstructed using the information from the silicon tracker (``inner tracks'') and the muon system (``outer tracks'')~\cite{Sirunyan:2018}.
The matching between inner and outer tracks is performed starting either from the tracks in the silicon trackers or from those reconstructed in the muon system.
To reconstruct very low \pt muons that do not traverse the entire detector, candidates where inner tracks are matched to segments in only one or two muon detector layers are also considered.
Muons are selected among the reconstructed muon track candidates by applying loose requirements on the track in the muon system and the inner tracker, taking into account also their compatibility with small energy deposits in the ECAL and HCAL.

Muons are required to have a relative isolation, $\mathcal{I}^{\PGm} < 0.35$, to discriminate between muons from \PZ boson decays and those originating from hadron decays within jets. Here $\mathcal{I}^{\PGm}$ is defined as:
	\begin{equation}
		\label{eqn:pfiso}
		{\mathcal I}^{\PGm} \equiv \Big( \sum \pt^\text{charged} + \max\big[ 0, \sum \pt^\text{neutral} +
		\sum \pt^{\Pgg} - \pt^{\PGm,\mathrm{PU}} \big] \Big) / \pt^{\PGm} ,
	\end{equation}
where $\sum \pt^\text{charged}$ is the scalar \pt sum of charged hadrons originating from the PV, whilst $\sum \pt^\text{neutral}$ and $\sum \pt^{\Pgg}$ are the scalar sums for neutral hadrons and photons, respectively.
The isolation requirement is defined using a cone of radius $\Delta R=0.3$ around the muon direction at the PV, with the angular distance between two particles $i$ and $j$ defined as $\Delta R(i,j) = \sqrt{\smash[b]{(\Delta\eta_{i,j})^{2} + (\Delta\phi_{i,j})^{2}}}$, where $\Delta\eta$ and $\Delta\phi$ are the differences in $\eta$ and azimuthal angle, respectively. The \pt sums are computed within this isolation cone. 
The quantity $\pt^{\PGm,\text{PU}}$ in Eq.~(\ref{eqn:pfiso}) is defined from the $\pt$ sum of all the charged hadrons $i$ not originating from the PV as $\pt^{\PGm,\text{PU}}\equiv0.5\sum_{i}\ensuremath{p_{\mathrm{T},i}^{\PGm,\mathrm{PU}}}\xspace$, where the factor of 0.5 corrects for using only the charged particles in the isolation cone~\cite{PUmitigationCMS}, thus accounting for the expected rate of charged and neutral hadrons. The $\pt^{\PGm,\text{PU}}$ contribution is subtracted in the definition of ${\mathcal I}^{\PGm}$ to correct for energy deposits arising from pileup interactions.

Electrons with $\pt^{\Pe} > 7\GeV$ within the geometrical acceptance of the detector, defined by $\abs{\eta^{\Pe}} < 2.5$~\cite{CMS:2020uim}, are reconstructed by combining the information from the ECAL and the tracker.
Their identification is carried out using a BDT algorithm sensitive to the presence of bremsstrahlung along the electron trajectory, the geometrical and momentum-energy matching with the corresponding cluster in the ECAL, the features of the electromagnetic shower in the ECAL, and observables that discriminate against electrons originating from photon conversions.
The isolation sums for electrons, defined similarly as for muons, are included in the BDT discriminant as well.
The inclusion of isolation sums improves the suppression of nonprompt electrons from hadron decays and from overlap of neutral and charged hadrons within jets~\cite{CMS:2020uim}, and has a better performance than a sequential analysis using the relative isolation. The improvement is of the order of 10-20\% depending on the $\pt$ and $\eta$ region.
The BDT for the electron identification and isolation is implemented using the \textsc{xgboost} library~\cite{xgboost} and the training is performed on a dedicated sample of simulated $\PZ$+jets events.
Events are divided into six mutually exclusive categories defined by two $\pt$ ranges ($7<\pt^{\Pe}<10\GeV$ and $\pt^{\Pe}>10\GeV$) and three $\eta$ selections corresponding to the central barrel ($\abs{\eta^{\Pe}}<0.8$), outer barrel ($0.800<\abs{\eta^{\Pe}}<1.479$), and endcaps ($1.479<\abs{\eta^{\Pe}}<2.500$).

Final-state radiation (FSR) photons from leptons are recovered with the following procedure. Photons reconstructed by the PF algorithm within $\abs{\eta^{\Pgg}}<2.4$ are considered as FSR objects if they have $\pt^{\Pgg} > 2\GeV$ and a relative isolation ${\mathcal I}^{\Pgg} <1.8$, where ${\mathcal I}^{\Pgg}$ is computed using PF charged hadrons, neutral hadrons, and photon candidates within a cone of $\Delta R = 0.3$.
Every FSR candidate is associated with the closest lepton in the event and is not retained if $\Delta R(\Pgg,\Pell)/(\pt^{\Pgg})^2>0.012\GeV^{-2}$ and $\Delta R(\Pgg,\Pell) > 0.5$.
For each lepton, the FSR candidate with the lowest value of $\Delta R(\Pgg,\Pell)/(\pt^{\Pgg})^2$, if any, is selected.
The photon candidates identified from the FSR recovery algorithm are excluded from the computation of the muon isolation.
FSR photons are included in invariant mass computations. 

The impact parameter significance is used to suppress nonprompt leptons from decays of hadrons or photon conversions. This variable is defined as the ratio of the 3-dimensional impact parameter (computed with respect to the position of the PV) to its uncertainty. Leptons are rejected if the value of this quantity is greater than 4.

The momentum scale and resolution of electrons and muons are calibrated in bins of $\pt^{\Pell}$ and $\eta^{\Pell}$ using the leptonic decays of known dilepton resonances, as described in Refs.~\cite{CMS:2020uim,Sirunyan:2018}.
A tag-and-probe technique~\cite{CMS:2011aa} is used to measure the efficiency of the lepton reconstruction and selection criteria in data and simulation, using samples of \PZ boson events. The efficiencies are measured in bins of $\pt^{\Pell}$ and $\eta^{\Pell}$, and the simulated yields are corrected by the ratio of the efficiencies measured in the data and in the simulation.

The selection targets events containing at least four well-identified and isolated leptons consistent with \HZZfl production. The event selection is described below, and closely follows that employed in Refs.~\cite{CMSHIG19001, CMSHIG21009}.

The \PZ boson candidates are built from pairs of same-flavor and opposite-sign leptons ($\Pep\Pem$, $\PGmp\PGmm$) with an invariant mass in the range $12 < \mll < 120\GeV$.
They are then combined to form $\PZ\PZ$ candidates. The \PZ boson candidate with the invariant mass closest to the world-average \PZ boson mass~\cite{ParticleDataGroup:2024cfk} is referred to as $\PZ_1$, whilst $\PZ_2$ denotes the other \PZ boson candidate.
The flavors of the four leptons in the event are used to define three mutually exclusive channels: $4\Pe$, $4\PGm$, and $2\Pe 2\PGm$.

In order to be considered in the analysis, the $\PZ\PZ$ candidates must satisfy additional requirements aimed at improving the sensitivity to Higgs boson decays.
The invariant mass of the $\PZ_1$ candidates is required to be larger than 40\GeV.
All lepton pairs $(\Pell_i, \Pell_j)$ must be separated by an angular distance $\Delta R(\Pell_i, \Pell_j) > 0.02$, in order to avoid overlaps.
All selected events must contain at least two leptons with $\pt > 10\GeV$ and at least one lepton with $\pt > 20\GeV$.
In the $4\Pe$ and $4\PGm$ channels, where the same four leptons can be used to build an alternative $\PZ_a \PZ_b$ candidate, candidates with $m_{\PZ_b}<12\GeV$ are not considered if $\PZ_a$ is closer to the nominal \PZ boson mass than $\PZ_1$, and in this case the event is rejected.
This requirement removes events with an on-shell \PZ boson accompanied by a low-mass dilepton resonance (\eg, $\cPJgy$ or $\Upsilon$).
To further suppress events with leptons originating from hadron decays in jet fragmentation or from leptonic decays of low-mass resonances, the invariant mass of the four possible opposite-charge lepton pairs (irrespective of flavor), computed without FSR photons, must satisfy the condition $m_{\Pell^{+}\Pell'^{-}} > 4\GeV$.
Finally, the $\PZ\PZ$ candidates are retained if the invariant mass of the four-lepton system, $\mllll$, is larger than $70\GeV$.

In events where more than one $\PZ\PZ$ candidate satisfies the selection requirements above, the one chosen is that with the largest scalar \pt sum of the two leptons defining the $\PZ_2$ candidate.
The signal region considered in the analysis is composed of all the events with one $\PZ\PZ$ candidate passing the selection and satisfying the condition $105 < \mllll < 160\GeV$.

\section{Background estimation}
\label{sec:bkgd}

\subsection{Irreducible backgrounds}
\label{sec:irrbkgd}

The irreducible background to the $\PH\to\PZ\PZ\to4\Pell$ signal consists of $\PZ\PZ$ production from quark-antiquark annihilation or gluon fusion and is estimated using simulated events.
Background contributions arising from triple vector boson production and from the production of top quarks associated with vector bosons are negligible in the signal region and thus not considered in this analysis.
The irreducible background contributions are included as binned templates in the likelihood function separately for each of the reconstructed final states ($4\Pe$, $4\PGm$, and $2\Pe 2\PGm$).
The templates are normalized to the most accurate theoretical calculations for the $\qqbar\to\PZ\PZ\to 4\Pell$ and $\Pg\Pg\to\PZ\PZ \to 4\Pell$ ($\Pell = \Pe, \PGm, \tau$) cross sections, as described in Section~\ref{sec:samples}.

\subsection{Reducible background}
\label{sec:redbkgd}
The reducible background to the $\PH\to\PZ\PZ\to4\Pell$ signal arises from processes in which decays of hadrons, or misidentified jets are incorrectly reconstructed as leptons.
This contribution, referred to as ``$\text{ZX}$'', is estimated using control samples in data, as described in Ref.~\cite{CMSHIG19001}.

The method is based on the lepton misidentification rate, defined as the probability of a nonprompt lepton to pass also the final selection criteria. The rate is estimated in a control sample that includes a \PZ boson and exactly one additional ``loose'' lepton ($\PZ$+$\Pell$); the latter fulfills the \pt, $\eta$, and PV requirements, but not those on identification and isolation.   
The lepton misidentification rate is then applied to two other control samples, consisting of a \PZ boson candidate and two opposite- or same-sign ``loose'' leptons  ($\PZ$+$\Pell\Pell$), to estimate the number of events in the signal region.

The reducible background contribution is included as a binned template in the likelihood function described in Section~\ref{sec:measurement} separately for the three considered final states ($4\Pe$, $4\PGm$, and $2\Pe 2\PGm$).

\section{Fiducial phase space definition}
\label{sec:fiducialvolume}
In order to reduce the impact of specific theoretical models on the acceptance, the cross sections are measured in a fiducial phase space defined to match closely the experimental acceptance defined by the reconstruction-level selections.
The fiducial phase space is defined at the generator level, following the strategy adopted in previous \HZZfl analyses~\cite{CMSH4l2016,CMSHIG19001}.
It is defined in terms of the lepton kinematics and isolation, and of the topology of the event.
The fiducial phase space definition is summarized in Table~\ref{tab:FidDef}.

Leptons are used at the generator level; they are obtained by combining the four-momentum of each lepton after photon FSR with that of the radiated photons found within a cone of radius $\Delta R = 0.3$ around the lepton.
The events are retained if the leading (subleading) lepton has $\pt>20\, (10)\GeV$. Additional electrons (muons) that may be present in the event are required to have  $\pt>7\,(5)\GeV$ and $\abs{\eta}<2.5\, (2.4)$.
The isolation of the leptons is ensured by requiring the scalar \pt sum of all stable particles, \ie, those particles not decaying in the detector volume, within a cone of radius $\Delta R=0.3$ to be less than 0.35 times the lepton \pt.
Neutrinos, FSR photons, and leptons (electrons and muons) are not included in the computation of the isolation sum to enhance the model independence of the measurements~\cite{CMSH4lFiducial8TeV}.
Events passing these requirements are retained if they have at least two same-flavor, opposite-sign lepton pairs.
The pair with invariant mass closest to the nominal \PZ boson mass~\cite{ParticleDataGroup:2024cfk} is labeled as $\PZ_{1}$ and must have $40<m_{\PZ_{1}}<120\GeV$.
The second \PZ boson candidate is referred to as $\PZ_{2}$ and must satisfy the requirement $12<m_{\PZ_{2}}<120\GeV$.
Each lepton pair $\Pell_i, \Pell_j$ must be separated by $\Delta R(\Pell_i, \Pell_j) > 0.02$, while any opposite-sign lepton pair must have $m_{\Pell^{+}\Pell'^{-}} > 4\GeV$, reflecting the selection criteria used at reconstruction level.
The signal region is defined by events satisfying the invariant mass requirement $105<m_{4\Pell}<160\GeV$.
Electrons and muons produced in the decay of a $\tau$ lepton are excluded from the definition of the fiducial region.
Nonfiducial events, \ie, events at the reconstruction level that do not satisfy the fiducial requirements, are treated as due to background processes in the fit described below.

\begin{table}[!htb]
	\begin{center}
		\caption{
			Summary of requirements used in the definition of the fiducial phase space for the $\Hllll$ cross section measurements.
			\label{tab:FidDef}
		}
		\begin{tabular}{lc}
			\multicolumn{2}{c}{Lepton kinematics and isolation} \\
			\hline 
			Leading lepton $\pt$ & $> 20\GeV$ \\
			Next-to-leading lepton $\pt$ & $> 10\GeV$ \\
			Additional electrons (muons) $\pt$ & $> 7(5)\GeV$ \\
			Pseudorapidity of electrons (muons) & $\abs{\eta} < 2.5(2.4)$ \\
			$\pt$ sum of all stable particles within $\Delta R < 0.3$ from each lepton & $< 0.35 \pt^\Pell$ \\[2ex]

			\multicolumn{2}{c}{Event topology} \\
			\hline 
			\multicolumn{2}{l}{at least two same-flavor, opposite-charge lepton pairs} \\
			Inv. mass of the Z$_1$ candidate & $40 < m($Z$_{1}$)$ < 120\GeV$ \\
			Inv. mass of the Z$_2$ candidate & $12 < m($Z$_{2}$)$ < 120\GeV$ \\
			Distance between selected four leptons & $\Delta R(\Pell_{i}\Pell_{j})>0.02$ for any $i\neq j$  \\
			Inv. mass of any opposite sign lepton pair & m$(\Pell^{+}\Pell'^{-})>4\GeV$ \\
			Inv. mass of the selected four leptons & $105 < m_{4\Pell} < 160\GeV$  \\
			\multicolumn{2}{l}{the selected four leptons must originate from the $\Hllll$ decay} \\
		\end{tabular}
	\end{center}
\end{table}

\section{Measurement strategy}
\label{sec:measurement}
The differential fiducial cross section for $\Pp\Pp\to \PH \to4\Pell$ is measured by performing an unbinned maximum likelihood fit of the signal and background contributions to the observed $4\Pell$ mass distribution, $N_{\mathrm{obs}}(m_{4\Pell})$, and the fiducial cross section ($\sigma_{\text{fid}}$) is directly extracted from the fit. The value of \mH is fixed to 125.38\GeV~\cite{CMS:2020xrn} in the fit.

The number of expected events in each final state f and in each bin $i$ of a given observable is expressed as a function of $m_{4\Pell}$ as:
\begin{eqnarray}
\label{eqn:m4l}
\begin{aligned}
N_{\text{obs}}^{\mathrm{f},i}(m_{4\Pell}) &= N_{\text{fid}}^{\mathrm{f},i}(m_{4\Pell})+N_{\text{nonfid}}^{\mathrm{f},i}(m_{4\Pell})+N_{\text{nonres}}^{\mathrm{f},i}(m_{4\Pell})+N_{\text{bkg}}^{\mathrm{f},i}(m_{4\Pell}) \\
&=\sum_j^{\text{genBin}}\epsilon_{i,j}^{\mathrm{f}} \left(1+f_\text{nonfid}^{\mathrm{f},i} \right) \sigma_{\text{fid}}^{\mathrm{f},j} \mathcal{L} \mathcal{P}_{\mathrm{res}}(m_{4\Pell}) \\
&~~~~~+ N_{\text{nonres}}^{\mathrm{f},i} \mathcal{P}_{\text{nonres}}(m_{4\Pell})+N_{\text{bkg}}^{\mathrm{f},i} \mathcal{P}_{\text{bkg}}(m_{4\Pell}).
\end{aligned}
\end{eqnarray}
The parameter $\sigma_{\text{fid}}^{\mathrm{f},j}$ is the signal cross section in bin $j$ of the fiducial phase space, defined at the generator-level (\text{genBin}). This is the result of the measurement.
The quantities $N_\text{fid}^{\mathrm{f},i}(m_{4\Pell})$ and $N_\text{nonfid}^{\mathrm{f},i}(m_{4\Pell})$ represent the resonant contributions originating from within and outside the fiducial volume, respectively.
The term $N_\text{nonres}^{\mathrm{f},i}(m_{4\Pell})$ represents the contribution from combinatorial association of the four leptons in events with more than four leptons, arising from WH, ZH, and $\ttH$ production modes where one of the leptons from the Higgs boson decay is lost or not selected; the component $N_\text{bkg}^{f,i}(m_{4\Pell})$ is the contribution from the reducible/irreducible background. The quantities $\mathcal{P}_\mathrm{res}(m_{4\Pell})$, $\mathcal{P}_\text{nonres}(m_{4\Pell})$, and $\mathcal{P}_\text{bkg}(m_{4\Pell})$ are the corresponding probability density functions (pdfs), assuming that the resonant fiducial signal and resonant nonfiducial signal have the same pdf.
The Higgs boson resonant signal distribution, $\mathcal{P}_{\mathrm{res}}(m_{4\Pell})$, is parameterized with a double-sided Crystal Ball (DCB) function~\cite{Gaiser:1982yw, Oreglia:1980cs, Skwarnicki:1986xj} around $\mH=125\GeV$.
The corresponding pdf, $\mathcal{P}_{\mathrm{res}}(m_{4\Pell})$, is scaled by the fiducial cross section, $\sigma_{\text{fid}}$, and the integrated luminosity $\mathcal{L}$.
The DCB function parameters are obtained from a simultaneous fit of the $\mllll$ distributions corresponding to the various mass points in the $\mH$ range 105--160\GeV, which allows expressing the dependence of the fitted parameters on $\mH$ directly in the fit, following the same strategy of Refs.~\cite{CMSHIG19001, CMSHIG21009}.

The shape of the nonresonant signal contribution, $\mathcal{P}_{\text{nonres}}(m_{4\Pell})$, is modeled by a Landau distribution with shape parameters constrained in the fit to be within a range determined from simulation. This contribution amounts to around 1\% of the total signal depending on the final state and it is treated as background; hereafter, it is referred to as ``nonresonant signal''.

An additional contribution ($f_\text{nonfid}^{\mathrm{f},i}$) is introduced, for each bin $i$ and final state f, to take into account events not originating from the fiducial volume but satisfying the selection criteria.
This contribution is referred to as the ``signal-induced background'' and is estimated from simulation for each production mode.
To minimize the model dependence of the measurement, the value of $f_\text{nonfid}^{\mathrm{f},i}$ is fixed to be a fraction of the fiducial signal component.

Generator-level observables used in the definition of the fiducial phase space are smeared by detector effects at reconstruction level.
The $\epsilon_{i,j}^{\mathrm{f}}$ response matrix is obtained from simulation, for each final state f, and is used to unfold the number of expected events in bin \textit{i} at the reconstruction level to the number of expected events of a given observable in bin \textit{j} at the fiducial level.
These efficiency numbers are obtained under the assumption that the ratios of the production modes yields are those predicted by the SM and include the per-lepton corrections and scale factors.
The kinematic acceptance is defined as the fraction of signal events that fall within the fiducial phase space.

Systematic uncertainties are included as nuisance parameters and the fiducial cross section measurements are obtained using an asymptotic approach~\cite{LHC-HCG} with a test statistic based on the profile likelihood ratio~\cite{Cowan_2011}.
A maximum likelihood fit is performed simultaneously in all final states and bins of each observable.
Two additional parameters regulating the mixture of the three different final states ($4\Pe,4\PGm,2\Pe2\PGm$) in the analysis are included in the fit and left floating to increase the model independence of the measurements, following the strategy adopted in Refs.~\cite{CMSHIG19001, CMSHIG21009}.
A likelihood-based unfolding is performed to resolve the detector effects from the observed distributions in the fiducial phase space.
This approach is the same as that described in Refs.~\cite{CMSHggFiducial8TeV,CMSHIG19001, CMSHIG21009, CMS:2020dvg, CMS:2020hkd} and allows the simultaneous unfolding of detector effects and the extraction of the fiducial cross section~\cite{CAT-23-001}.

\section{Systematic uncertainties}
\label{sec:systematics}
The systematic effects considered in this analysis closely reflect those studied in Ref.~\cite{CMSHIG21009}. All systematic uncertainties, which are modeled with nuisance parameters, have been re-assessed using the 2022 data set, except for those related to $K$ factors used in the modeling of irreducible background processes.

The integrated luminosity of the 2022 data-taking period considered in the analysis is known with an uncertainty of 1.4\%~\cite{CMS-PAS-LUM-22-001}.

Experimental systematic uncertainties due to the trigger, and lepton reconstruction and selection efficiencies are estimated from data for the different final states. These uncertainties are derived from a tag-and-probe technique using $J/\psi$ and Z decays into a pair of leptons. 
They range from 0.8 to 1.8\% in the 4\PGm channel and from 6 to 11\% in the 4\Pe channel.
The difference between the two final states reflects the use of $J/\psi \to \PGm\PGm$ events for the estimate of the muon uncertainties at low $\pt^{\PGm}$, whereas only the Z boson resonance is utilized for the electron case, leading to larger uncertainties in the low-$\pt^{\Pe}$ region.

The systematic uncertainties in the lepton momentum scale and resolution are estimated from dedicated studies on the $Z\rightarrow \Pell^+\Pell^-$ mass distribution in data and simulation.
Specifically, the uncertainty is assessed by propagating to the four-lepton invariant mass the uncertainty associated to the momentum corrections for each individual lepton.
The four-lepton mass distributions obtained by varying the four-lepton momentum up or down by one standard deviation are fitted, with only the mean value floating, and the difference between the new mean value and the nominal one is taken as an estimate of the sensitivity to the lepton energy scale.
In the 4\Pe channel, the scale uncertainty is found to be 0.2\%, while the resolution uncertainty is 12\%.
In the 4\PGm channel the scale uncertainty is 0.05\% and the resolution uncertainty 5\%.
In the mixed flavor \tetmu channel, the scale and the resolution uncertainties are 0.13\% and 8.5\%, respectively.
The effect of these uncertainties is introduced in the analysis by using additional floating factors added to the corresponding parameters of the DCB function used to model the resonant signal.
These uncertainties are not applied to the $\PZ\PZ$ backgrounds estimated from simulation since their distribution is almost flat under the \mH~peak; therefore, shifts in the scale have a negligible effect.

The following systematic effects in the reducible ZX background determination are studied: the statistical uncertainty in the number of events in the control regions, the effect of $\pm1$ standard deviation variations of the misidentification rates, and the difference in composition among various processes that contribute to this background. The overall effect of these three sources ranges between 25\% and 46\%, depending on the final state, and is included as a nuisance parameter with a log-normal prior in the fit.

The systematic uncertainty in the $K$ factors used in the modeling of the irreducible background processes is also considered. A 10\% uncertainty is assumed for the $K$ factor used in the \ggZZ prediction, while a 0.1\% average uncertainty affects the $K$ factor for the $\qqZZ$ electroweak corrections. These uncertainties are derived from the theory predictions of the $K$ factors~\cite{CMSHIG21009}. 

Theoretical uncertainties in the renormalization and factorization scales and PDFs used for the estimate of the irreducible backgrounds are studied, as they may affect the rates of these processes. The uncertainty from the renormalization and factorization scales is determined by varying them between 0.5 and 2 times their nominal value, while keeping their ratio between 0.5 and 2; this yields an overall 4\% effect. The uncertainty due to PDFs is determined following the PDF4LHC recommendations by taking the root mean square of the variation of the results when using different replicas of the default NNPDF set. The effect is around 3\% for both irreducible background processes considered. The effect of the renormalization and factorization scale uncertainties range between 0.3\% and 6\% across the different production mechanisms, while the effect of the PDF uncertainties is between 1.6\% and 3.6\%. The effect of theoretical uncertainties on the signal is negligible for the present analysis and is not included in the fit.

\section{Results}
\label{sec:results}
\begin{figure}[!tbp]
    \centering
    {\includegraphics[width=0.48\textwidth]{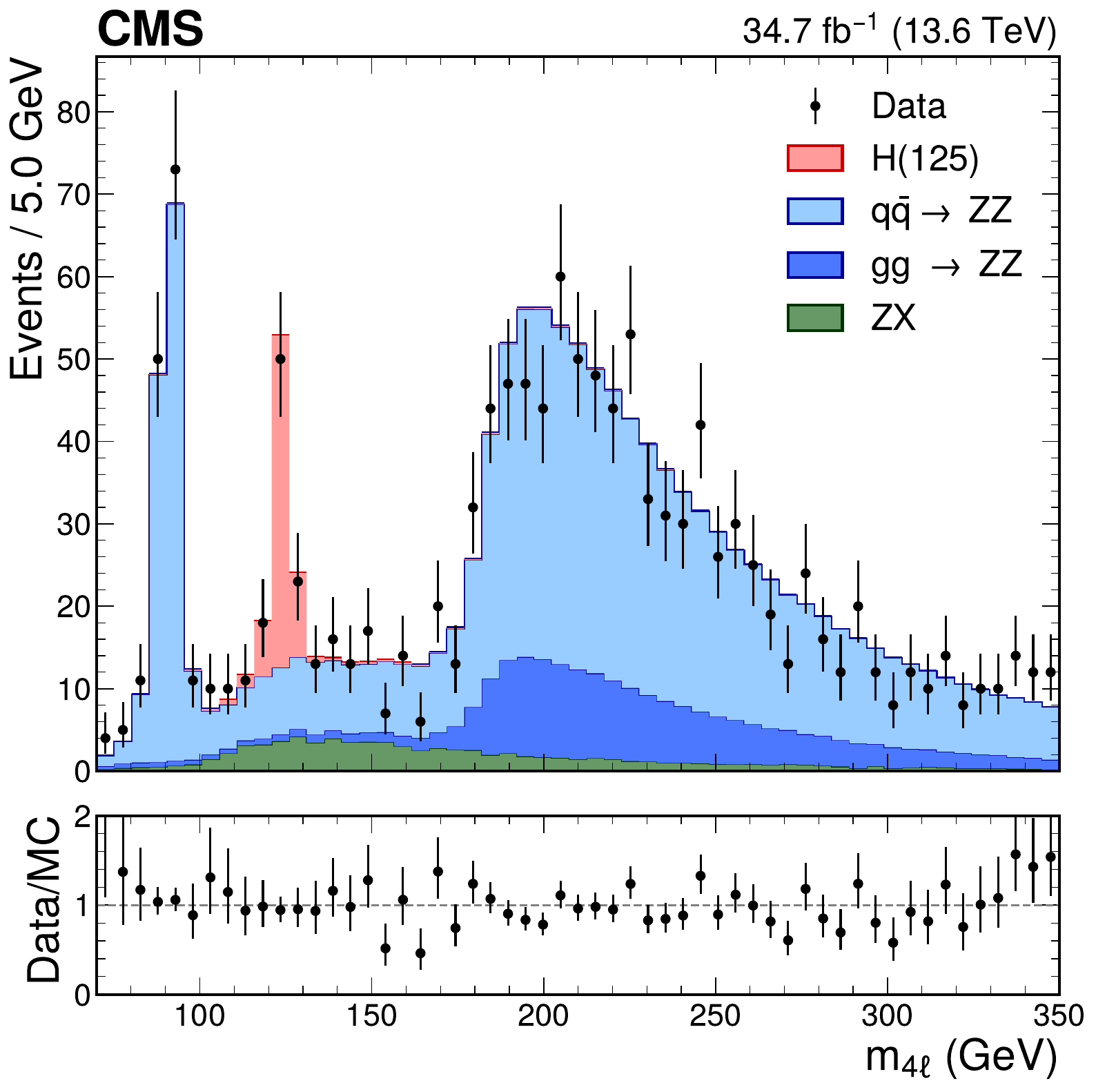}}
	{\includegraphics[width=0.48\textwidth]{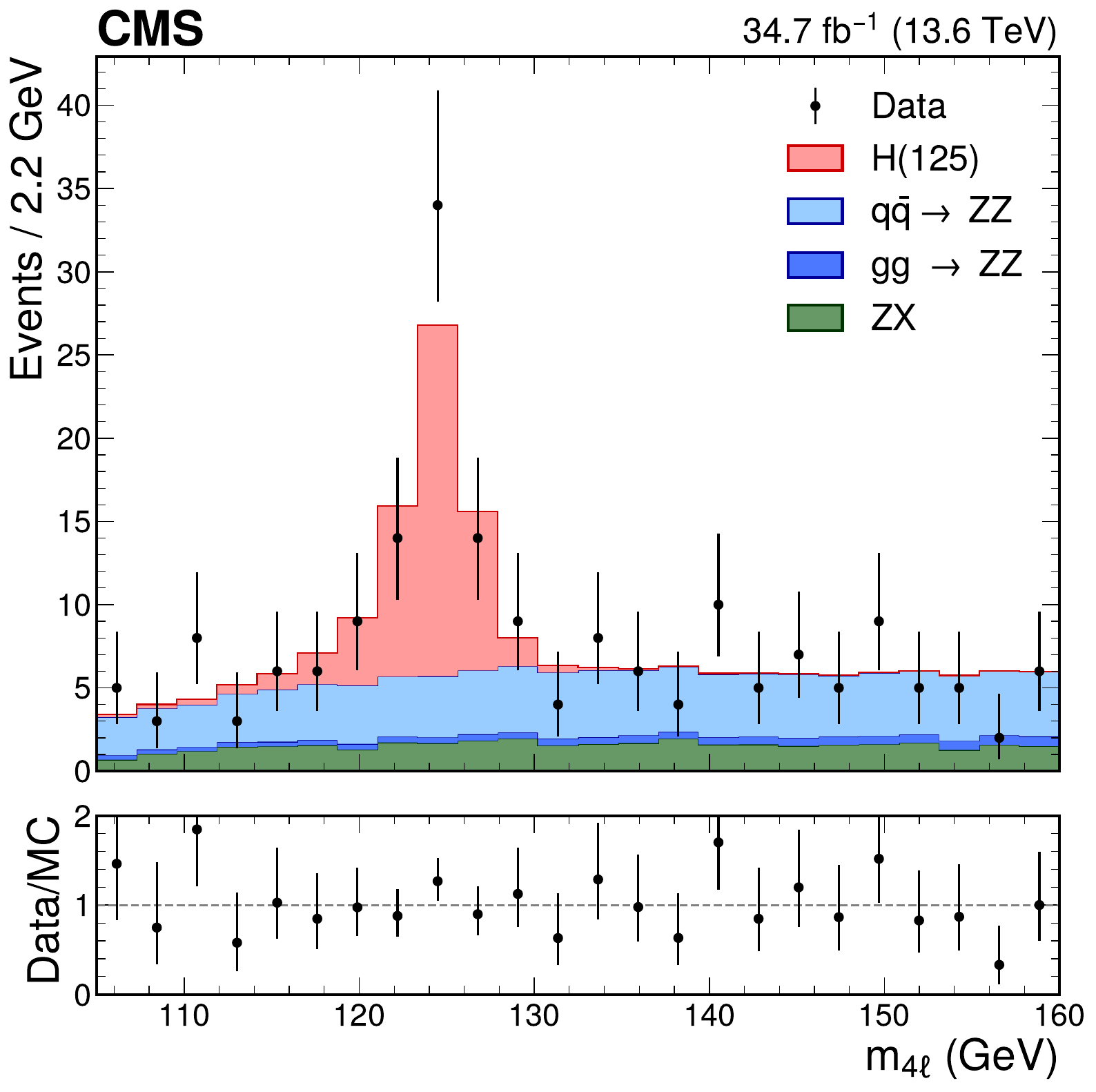}}
    \caption{Distribution of the $\mllll$ invariant mass in the ranges 70--350\GeV (left), and 105--160\GeV (right). The black points with error bars represent the data. The colored histograms show the signal (red histogram) and the background contributions. The post-fit normalization for all the processes is obtained from the measurement performed in the range $105<\mllll<160\GeV$ and then ported to all other distributions and ranges by scaling it by the post-fit/pre-fit yield ratio.
    The bottom panels depict the ratio of the data to the post-fit MC distribution.}
    \label{fig:m4l_tot}
\end{figure}

The results are obtained from a simultaneous fit of Eq.~(\ref{eqn:m4l}) to the data for the three final states (\fe, \fmu, and \tetmu).
The numbers of expected events after the fit to data (post-fit) in the mass range $105<\mllll<160\GeV$, for each final state and an integrated luminosity of \usedLumiRunIII, are shown in Table~\ref{tab:postfityields}.
The $\mllll$ invariant mass distribution for the inclusive $4\Pell$ final state is presented in Fig.~\ref{fig:m4l_tot}.
Tabulated results are provided in the HEPData record for this analysis~\cite{hepdata}.

\begingroup
\renewcommand{\arraystretch}{1.5}
\begin{table}[t]
	\begin{center}
		\caption{Post-fit yields for the four final states in the signal region ($105<\mllll<160\GeV$).
                The ``nonfid'' contribution arises from signal events not originating from the fiducial volume but satisfying the analysis selection, while the ``nonres'' contribution contains signal events from VH or $\ttH$ where one of the leptons from the Higgs boson decay is lost or not selected (details in Section~\ref{sec:fiducialvolume}). The contributions of signal, nonfid, and nonres events are estimated assuming $m_{\PH} = 125.38\GeV$.}
			\label{tab:postfityields}

			\begin{tabular}{cccccc}
			Process & $4\Pe$ & $4\PGm$ & $2\Pe2\PGm$ & $4\Pell$   \\
			\hline 
Signal                      & $10.69^{+0.98}_{-1.10}$ & $12.09^{+0.20}_{-0.29}$ & $29.9^{+1.9}_{-2.0}$    & $52.6^{+2.9}_{-3.1}$     \\
Nonfid                      & $0.38^{+0.04}_{-0.04}$  & $0.28^{+0.01}_{-0.01}$  & $0.35^{+0.02}_{-0.02}$  & $1.01^{+0.06}_{-0.07}$   \\
Nonres                      & $0.11^{+0.01}_{-0.01}$  & $0.22^{+0.01}_{-0.01}$  & $0.33^{+0.02}_{-0.02}$  & $0.66^{+0.03}_{-0.04}$   \\
Total signal                & $11.2^{+1.0}_{-1.1}$    & $12.59^{+0.21}_{-0.30}$ & $30.5^{+1.9}_{-2.0}$    & $54.3^{+3.0}_{-3.2}$     \\
\hline
qqZZ                        & $13.4^{+1.4}_{-1.6}$    & $33.2^{+1.6}_{-2.0}$    & $39.0^{+3.0}_{-3.3}$    & $85.6^{+5.4}_{-6.3}$     \\
ggZZ			    & $1.92^{+0.28}_{-0.28}$  & $3.93^{+0.45}_{-0.42}$  & $4.00^{+0.51}_{-0.49}$  & $9.8^{+1.2}_{-1.1}$      \\
ZX			    & $4.4^{+1.3}_{-1.5}$     & $15.0^{+3.9}_{-3.6}$    & $18.0^{+3.5}_{-3.4}$    & $37.5^{+5.4}_{-5.1}$     \\
Sum of backgrounds 	    & $19.7^{+2.1}_{-2.3}$    & $52.2^{+4.3}_{-4.1}$    & $61.0^{+4.8}_{-4.9}$    & $132.9^{+8.0}_{-8.4}$    \\
\hline
Total expected		    & $30.9^{+2.9}_{-3.2}$    & $64.8^{+4.3}_{-4.2}$    & $91.5^{+6.1}_{-6.3}$    & $187.2^{+9.9}_{-10.5}$   \\
\hline
Total observed		    & $32$                    & $59$                    & $93$                    & $184$                    \\
\end{tabular}
	\end{center}
\end{table}
\endgroup

\begin{figure}[!tbp]
\centering
{\includegraphics[width=0.55\textwidth]{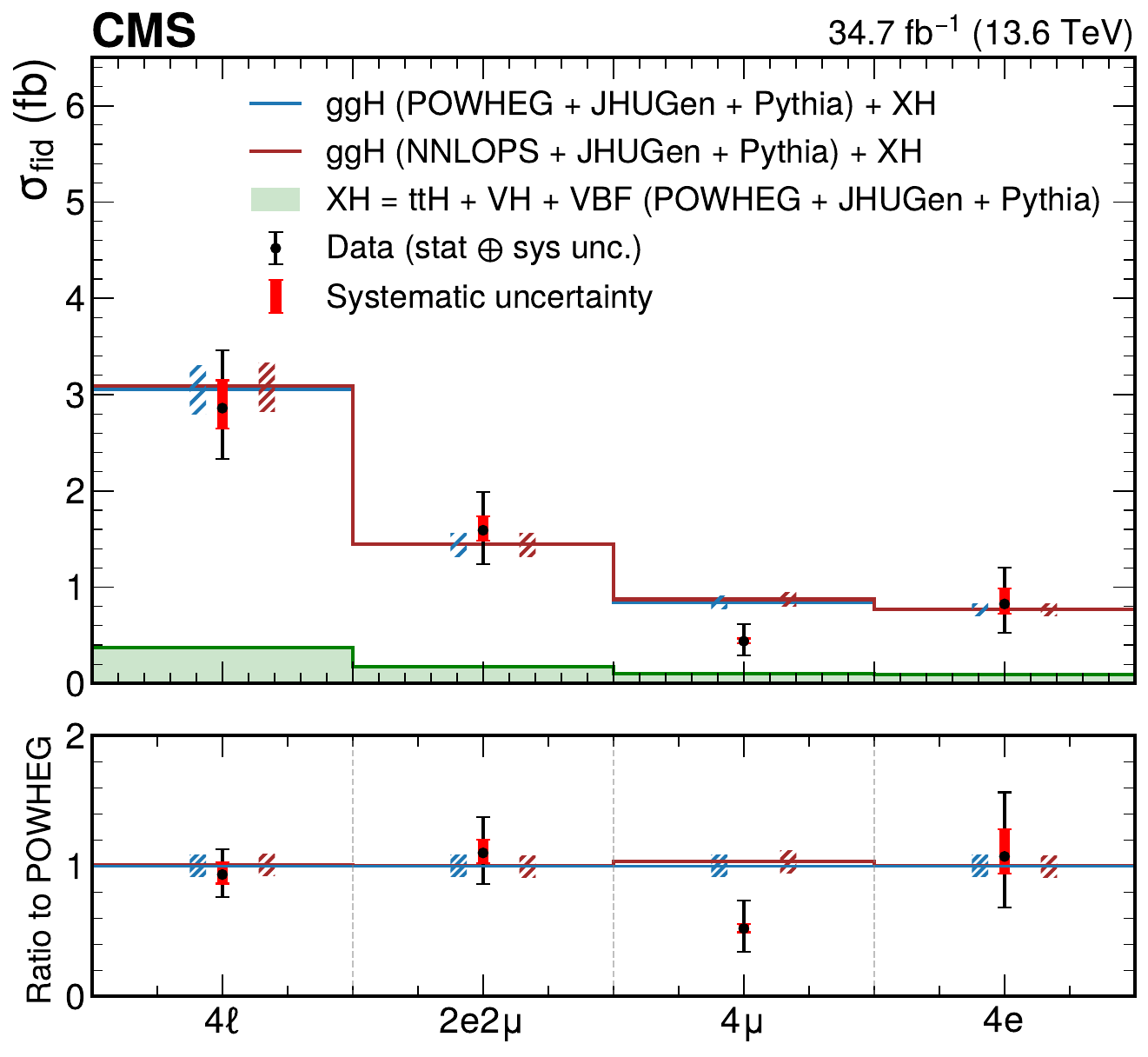}}
\caption{Measured inclusive fiducial \HZZfl cross section in the various final states at 13.6\TeV.
In the upper plot, the acceptance and theoretical uncertainties in the differential bins are calculated using the $\ggH$ predictions from two different generators normalized to next-to-NNLO order. The subdominant component of the signal ($\text{VBF} + \text{VH} + \ttH$) is denoted as XH and is fixed to the SM prediction. The measured cross sections are compared with the  $\ggH$ predictions from \POWHEG (blue) and \textsc{NNLOPS} (dark red). The hatched areas correspond to the systematic uncertainties of the theoretical predictions. Black points represent the measured fiducial cross sections in each bin, black error bars the total uncertainty of each measurement, and red boxes the systematic uncertainties.
The lower panel displays the ratio of the measured cross sections to the \POWHEG predictions, as well as the ratio of the \textsc{NNLOPS} predictions to those from \POWHEG.
}
\label{fig:nll_scan_sqrts}
\end{figure}

\begin{figure}[!tbp]
	\centering
	\includegraphics[width=0.65\textwidth]{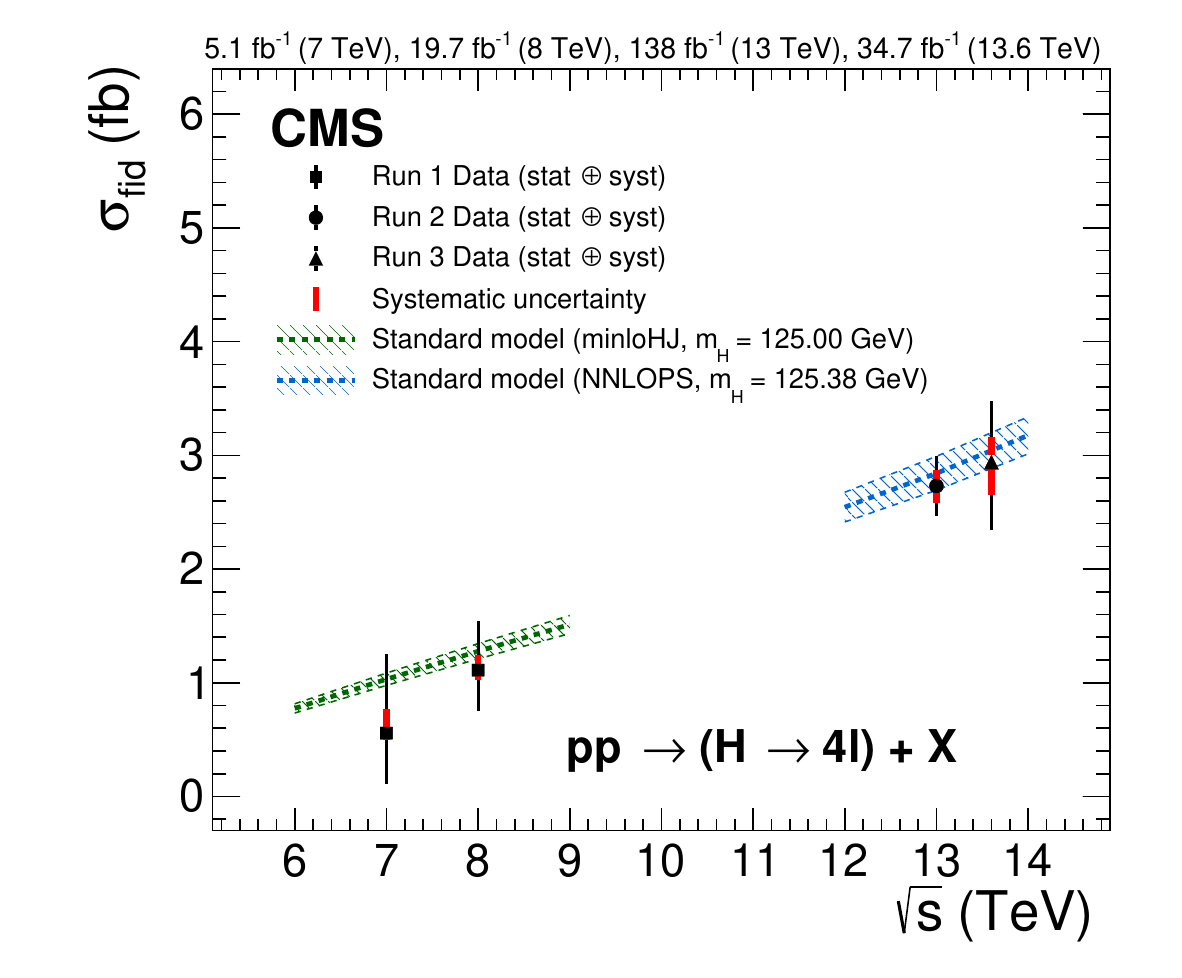}
	\caption{Measured inclusive fiducial \HZZfl cross section as a function of the center-of-mass energy $\sqrt{s}$. The acceptance is calculated using \textsc{minloHJ}~\cite{minlo} at $\sqrt{s}=$~7 and 8\TeV and \textsc{NNLOPS}~\cite{Hamilton:2013fea} at $\sqrt{s}=13$ and 13.6\TeV.
	}
	\label{fig:nll_scan_sqrts_1}
	\end{figure}

The measured inclusive fiducial cross section for the \HZZfl process at $\mH = 125.38\GeV$ is found to be:
\begin{equation}
\begin{split}
\sigma_\text{fid} & = 2.89^{+0.53}_{-0.49}\stat^{+0.29}_{-0.21}\syst\fb \\
                    & = 2.89^{+ 0.53}_{-0.49}\stat^{+0.26}_{-0.19}\,\text{(electrons)}^{+0.06}_{-0.06}\,\text{(ZX)}^{+0.06}_{-0.05}\lum^{+0.04}_{-0.04}\,\text{(bkg)} ^{+0.04}_{-0.03}\,\text{(muons)}\fb,
\end{split}
\end{equation}
in good agreement with the SM expectation of $3.09^{+0.27}_{-0.24}$\unit{fb}. The SM expectation is obtained by multiplying the cross section and branching fraction values for the process studied, taken from~\cite{deFlorian:2016spz}, by the acceptance. The uncertainties include contributions from factorization and renormalization scales, variations of PDF and strong coupling constant, and branching fractions.
The dominant source of systematic uncertainty on the measured cross section is the electron selection efficiency. 
Figure~\ref{fig:nll_scan_sqrts} shows the measured inclusive cross section in the three final states, while Fig.~\ref{fig:nll_scan_sqrts_1} shows the evolution of the \HZZfl fiducial cross section as a function of the center of mass energy.
Table~\ref{tab:fiducial_inclusive} summarizes the results for each of the final states studied.

\begin{table*}[tb]
	\begin{center}
	\topcaption{
                Measured fiducial cross sections for different final states for $\mH=125.38\GeV$.
		\label{tab:fiducial_inclusive}
	}
	\renewcommand{\arraystretch}{1.5}
	\begin{tabular}{lc}

		                  & $\sigma_\text{fid}$ (fb) \\
		\hline
		$2\Pe 2\PGm$ & $1.60^{+0.37}_{-0.33}\stat^{+0.14}_{-0.11}\syst$ \\
		$4\PGm$ & $0.46^{+0.18}_{-0.15}\stat^{+0.03}_{-0.03}\syst$ \\
		$4\Pe$  & $0.83^{+0.34}_{-0.29}\stat^{+0.16}_{-0.10}\syst$ \\
		Inclusive & $2.89^{+0.53}_{-0.49}\stat^{+0.29}_{-0.21}\syst$ \\

	\end{tabular}
	\end{center}
\end{table*}

\subsection{Differential cross section measurements}
\label{sec:diff_res}
\begin{figure}[!tbp]
    \centering
    {\includegraphics[width=0.48\textwidth]{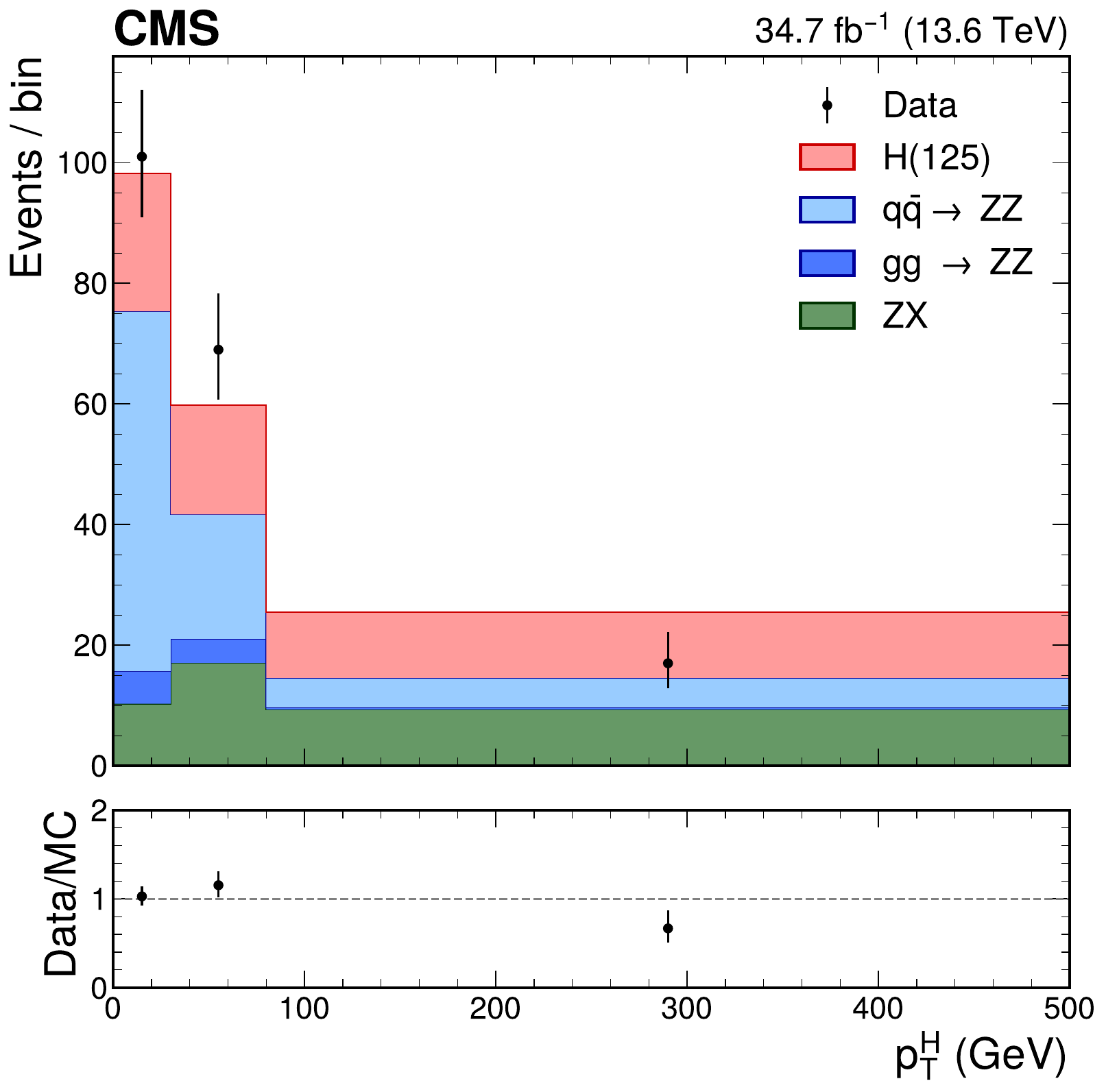}}
	{\includegraphics[width=0.48\textwidth]{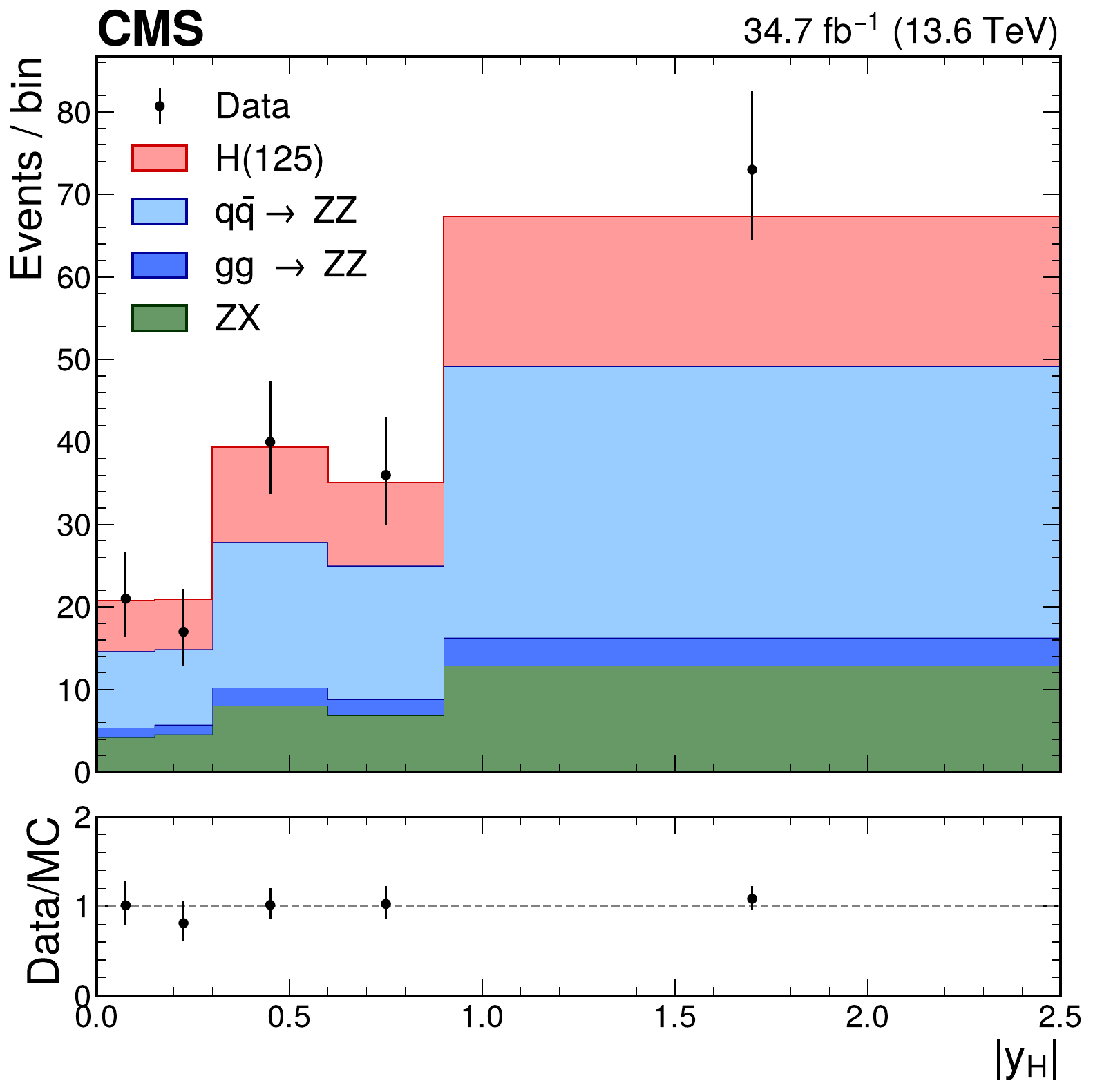}}
    \caption{Distributions of $\pt^{\PH}$ (left) and $\abs{y_{\PH}}$ (right) for events with $105<\mllll<160\GeV$. The black points with error bars represent the data. The colored histograms indicate the signal (red histogram) and the background contributions. The histograms are normalized to the post-fit yields reported in Table~\ref{tab:postfityields}.
    The bottom panels depict the ratio of the data to the post-fit MC distribution.}
    \label{fig:recolevel_pt_rapidity}
\end{figure}

\begin{figure}[!tbp]
    \centering
    {\includegraphics[width=0.6\textwidth]{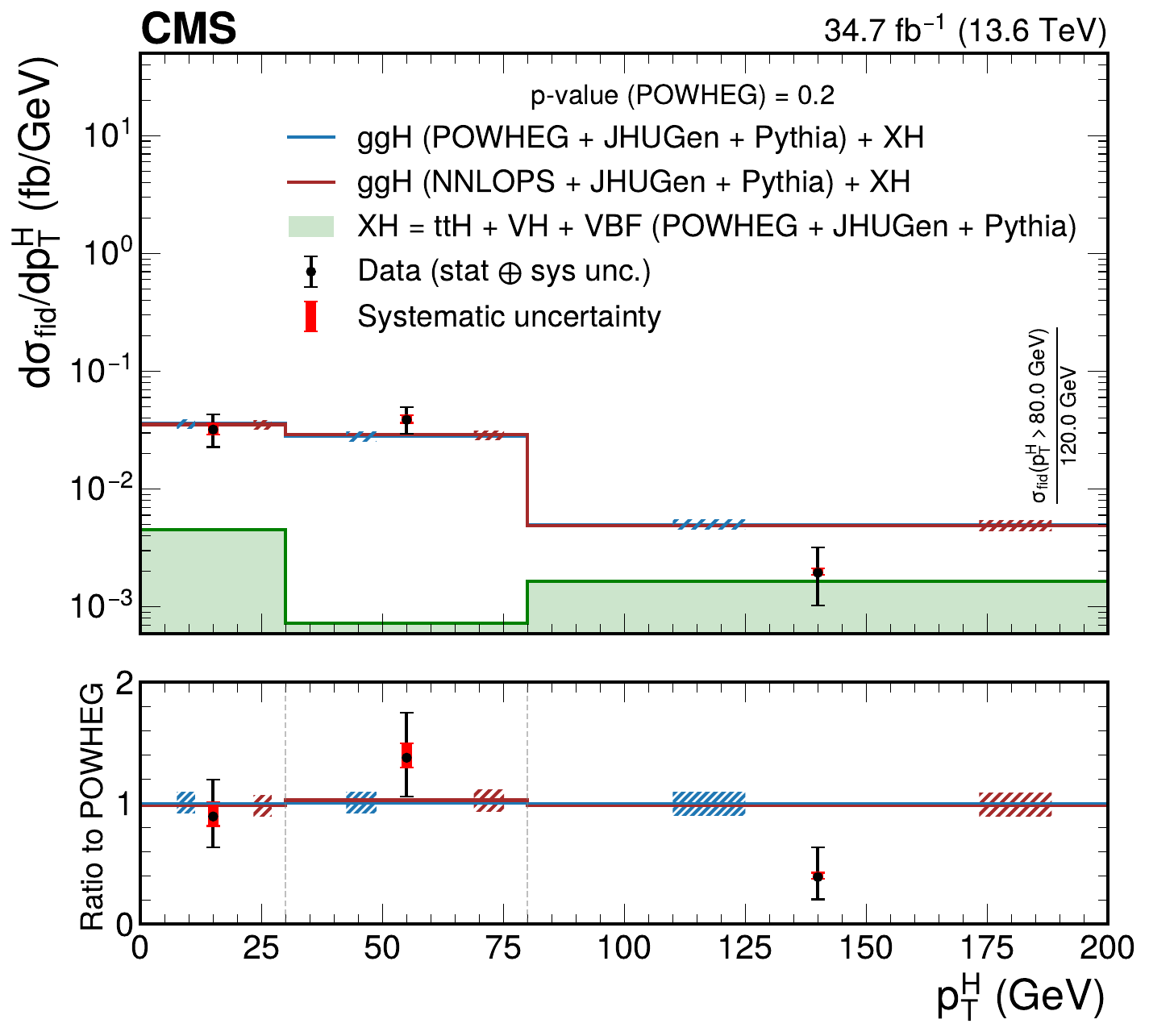}}
    {\includegraphics[width=0.6\textwidth]{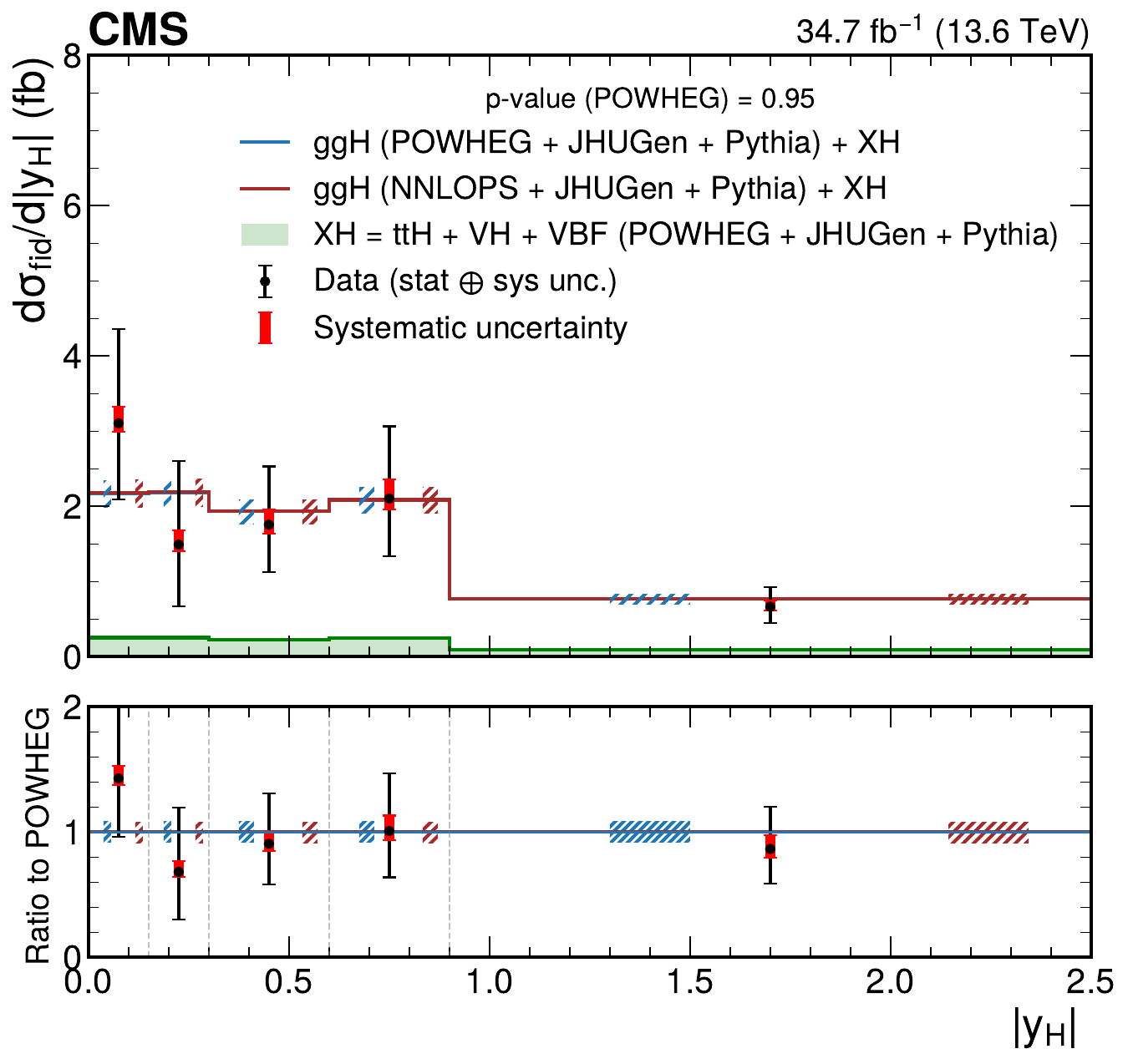}}
    \caption{Differential fiducial cross sections measured in the \HZZfl channel at 13.6\TeV as a function of $\pt^{\PH}$ (upper) and $\abs{y_{\PH}}$ (lower).
    The acceptance and theoretical uncertainties in the differential bins are calculated using the $\ggH$ predictions from two different generators normalized to next-to-NNLO order.
    The subdominant component of the signal ($\text{VBF} + \text{VH} + \ttH$) is denoted as XH and is fixed to the SM prediction.
    The measured cross sections are compared with the  $\ggH$ predictions from \POWHEG (blue) and \textsc{NNLOPS} (dark red).
    The hatched areas correspond to the systematic uncertainties of the theoretical predictions.
    The black points represent the measured fiducial cross sections in each bin, the black error bars the total uncertainties, and red boxes the systematic uncertainties.
    The fiducial cross section in the last bin of the upper plot is measured for events with $\pt^{\PH} > 80\GeV$ and normalized to a bin width of 120\GeV.
    The lower panels display the ratio of the measured cross sections to the \POWHEG predictions, as well as the ratio of the \textsc{NNLOPS} predictions to those from \POWHEG.
    The p-value is used to assess the compatibility of the results with the theoretical predictions and is found to be 0.2 and 0.95 for $\pt^{\PH}$ and $\abs{y_{\PH}}$, respectively.}
    \label{fig:diff_pt_rapidity}
\end{figure}

The fiducial cross section of the \HZZfl process is also measured differentially in bins of $\pt^{\PH}$ and $\abs{y_{\PH}}$.
The distribution of these two variables obtained after the full analysis selection is shown in Fig.~\ref{fig:recolevel_pt_rapidity} for the inclusive $4\Pell$ final state.
To perform the differential production cross section measurements, an unbinned maximum likelihood fit to the four-lepton invariant mass is performed in bins of $\pt^{\PH}$ and $\abs{y_{\PH}}$.
The distributions of the expected and observed cross sections in bins of $\pt^{\PH}$ are shown in the upper plot of Fig.~\ref{fig:diff_pt_rapidity}, while the lower plot shows the measurement as a function of $\abs{y_{\PH}}$.
In general, good agreement between the measured and predicted values is observed.

\section{Summary}
\label{sec:summary}
This paper presents the measurement of the fiducial production cross section of the Higgs boson (\PH) in the $4\Pell$ ($\Pell=\Pe,\PGm$) final state using the data collected with the CMS detector in 2022 at a center-of-mass energy of $\sqrt{s}=13.6\TeV$. The \HZZfl inclusive fiducial cross section is measured to be $2.89^{+0.53}_{-0.49}\stat^{+0.29}_{-0.21}\syst\unit{fb}$, in agreement with the standard model expectation of $3.09^{+0.27}_{-0.24}$\unit{fb}. The differential fiducial cross section is also measured in bins of transverse momentum and absolute value of the rapidity of the Higgs boson. All results are consistent with the standard model expectation.

\begin{acknowledgments}
We congratulate our colleagues in the CERN accelerator departments for the excellent performance of the LHC and thank the technical and administrative staffs at CERN and at other CMS institutes for their contributions to the success of the CMS effort. In addition, we gratefully acknowledge the computing centers and personnel of the Worldwide LHC Computing Grid and other centers for delivering so effectively the computing infrastructure essential to our analyses. Finally, we acknowledge the enduring support for the construction and operation of the LHC, the CMS detector, and the supporting computing infrastructure provided by the following funding agencies: SC (Armenia), BMBWF and FWF (Austria); FNRS and FWO (Belgium); CNPq, CAPES, FAPERJ, FAPERGS, and FAPESP (Brazil); MES and BNSF (Bulgaria); CERN; CAS, MoST, and NSFC (China); MINCIENCIAS (Colombia); MSES and CSF (Croatia); RIF (Cyprus); SENESCYT (Ecuador); ERC PRG, RVTT3 and MoER TK202 (Estonia); Academy of Finland, MEC, and HIP (Finland); CEA and CNRS/IN2P3 (France); SRNSF (Georgia); BMBF, DFG, and HGF (Germany); GSRI (Greece); NKFIH (Hungary); DAE and DST (India); IPM (Iran); SFI (Ireland); INFN (Italy); MSIP and NRF (Republic of Korea); MES (Latvia); LMTLT (Lithuania); MOE and UM (Malaysia); BUAP, CINVESTAV, CONACYT, LNS, SEP, and UASLP-FAI (Mexico); MOS (Montenegro); MBIE (New Zealand); PAEC (Pakistan); MES and NSC (Poland); FCT (Portugal); MESTD (Serbia); MCIN/AEI and PCTI (Spain); MOSTR (Sri Lanka); Swiss Funding Agencies (Switzerland); MST (Taipei); MHESI and NSTDA (Thailand); TUBITAK and TENMAK (Turkey); NASU (Ukraine); STFC (United Kingdom); DOE and NSF (USA).

\hyphenation{Rachada-pisek} Individuals have received support from the Marie-Curie program and the European Research Council and Horizon 2020 Grant, contract Nos.\ 675440, 724704, 752730, 758316, 765710, 824093, 899987, 101115353, 101002207, and COST Action CA16108 (European Union); the Leventis Foundation; the Alfred P.\ Sloan Foundation; the Alexander von Humboldt Foundation; the Science Committee, project no. 22rl-037 (Armenia); the Fonds pour la Formation \`a la Recherche dans l'Industrie et dans l'Agriculture (FRIA-Belgium); the Beijing Municipal Science \& Technology Commission, No. Z191100007219010 and Fundamental Research Funds for the Central Universities (China); the Ministry of Education, Youth and Sports (MEYS) of the Czech Republic; the Shota Rustaveli National Science Foundation, grant FR-22-985 (Georgia); the Deutsche Forschungsgemeinschaft (DFG), among others, under Germany's Excellence Strategy -- EXC 2121 ``Quantum Universe" -- 390833306, and under project number 400140256 - GRK2497; the Hellenic Foundation for Research and Innovation (HFRI), Project Number 2288 (Greece); the Hungarian Academy of Sciences, the New National Excellence Program - \'UNKP, the NKFIH research grants K 131991, K 133046, K 138136, K 143460, K 143477, K 146913, K 146914, K 147048, 2020-2.2.1-ED-2021-00181, TKP2021-NKTA-64, and 2021-4.1.2-NEMZ\_KI-2024-00036 (Hungary); the Council of Science and Industrial Research, India; ICSC -- National Research Center for High Performance Computing, Big Data and Quantum Computing and FAIR -- Future Artificial Intelligence Research, funded by the NextGenerationEU program (Italy); the Latvian Council of Science; the Ministry of Education and Science, project no. 2022/WK/14, and the National Science Center, contracts Opus 2021/41/B/ST2/01369 and 2021/43/B/ST2/01552 (Poland); the Funda\c{c}\~ao para a Ci\^encia e a Tecnologia, grant CEECIND/01334/2018 (Portugal); the National Priorities Research Program by Qatar National Research Fund; MCIN/AEI/10.13039/501100011033, ERDF ``a way of making Europe", and the Programa Estatal de Fomento de la Investigaci{\'o}n Cient{\'i}fica y T{\'e}cnica de Excelencia Mar\'{\i}a de Maeztu, grant MDM-2017-0765 and Programa Severo Ochoa del Principado de Asturias (Spain); the Chulalongkorn Academic into Its 2nd Century Project Advancement Project, and the National Science, Research and Innovation Fund via the Program Management Unit for Human Resources \& Institutional Development, Research and Innovation, grant B39G670016 (Thailand); the Kavli Foundation; the Nvidia Corporation; the SuperMicro Corporation; the Welch Foundation, contract C-1845; and the Weston Havens Foundation (USA).
\end{acknowledgments}\section*{Data availability} Release and preservation of data used by the CMS Collaboration as the basis for publications is guided by the  \href{https://doi.org/10.7483/OPENDATA.CMS.1BNU.8V1W}{CMS data preservation, re-use and open access policy}.

\bibliography{auto_generated}
\cleardoublepage \appendix\section{The CMS Collaboration \label{app:collab}}\begin{sloppypar}\hyphenpenalty=5000\widowpenalty=500\clubpenalty=5000
\cmsinstitute{Yerevan Physics Institute, Yerevan, Armenia}
{\tolerance=6000
V.~Chekhovsky, A.~Hayrapetyan, V.~Makarenko\cmsorcid{0000-0002-8406-8605}, A.~Tumasyan\cmsAuthorMark{1}\cmsorcid{0009-0000-0684-6742}
\par}
\cmsinstitute{Institut f\"{u}r Hochenergiephysik, Vienna, Austria}
{\tolerance=6000
W.~Adam\cmsorcid{0000-0001-9099-4341}, J.W.~Andrejkovic, L.~Benato\cmsorcid{0000-0001-5135-7489}, T.~Bergauer\cmsorcid{0000-0002-5786-0293}, S.~Chatterjee\cmsorcid{0000-0003-2660-0349}, K.~Damanakis\cmsorcid{0000-0001-5389-2872}, M.~Dragicevic\cmsorcid{0000-0003-1967-6783}, P.S.~Hussain\cmsorcid{0000-0002-4825-5278}, M.~Jeitler\cmsAuthorMark{2}\cmsorcid{0000-0002-5141-9560}, N.~Krammer\cmsorcid{0000-0002-0548-0985}, A.~Li\cmsorcid{0000-0002-4547-116X}, D.~Liko\cmsorcid{0000-0002-3380-473X}, I.~Mikulec\cmsorcid{0000-0003-0385-2746}, J.~Schieck\cmsAuthorMark{2}\cmsorcid{0000-0002-1058-8093}, R.~Sch\"{o}fbeck\cmsAuthorMark{2}\cmsorcid{0000-0002-2332-8784}, D.~Schwarz\cmsorcid{0000-0002-3821-7331}, M.~Sonawane\cmsorcid{0000-0003-0510-7010}, W.~Waltenberger\cmsorcid{0000-0002-6215-7228}, C.-E.~Wulz\cmsAuthorMark{2}\cmsorcid{0000-0001-9226-5812}
% ...
\par}
\vskip\cmsinstskip
\dag:~Deceased\\
$^{1}$Also at Yerevan State University, Yerevan, Armenia\\
$^{2}$Also at TU Wien, Vienna, Austria\\
$^{3}$Also at Ghent University, Ghent, Belgium\\
$^{4}$Also at Universidade do Estado do Rio de Janeiro, Rio de Janeiro, Brazil\\
$^{5}$Also at FACAMP - Faculdades de Campinas, Sao Paulo, Brazil\\
$^{6}$Also at Universidade Estadual de Campinas, Campinas, Brazil\\
$^{7}$Also at Federal University of Rio Grande do Sul, Porto Alegre, Brazil\\
$^{8}$Also at University of Chinese Academy of Sciences, Beijing, China\\
$^{9}$Also at China Center of Advanced Science and Technology, Beijing, China\\
$^{10}$Also at University of Chinese Academy of Sciences, Beijing, China\\
$^{11}$Also at China Spallation Neutron Source, Guangdong, China\\
$^{12}$Now at Henan Normal University, Xinxiang, China\\
$^{13}$Also at University of Shanghai for Science and Technology, Shanghai, China\\
$^{14}$Now at The University of Iowa, Iowa City, Iowa, USA\\
$^{15}$Also at an institute formerly covered by a cooperation agreement with CERN\\
$^{16}$Also at Helwan University, Cairo, Egypt\\
$^{17}$Now at Zewail City of Science and Technology, Zewail, Egypt\\
$^{18}$Now at British University in Egypt, Cairo, Egypt\\
$^{19}$Now at Cairo University, Cairo, Egypt\\
$^{20}$Also at Purdue University, West Lafayette, Indiana, USA\\
$^{21}$Also at Universit\'{e} de Haute Alsace, Mulhouse, France\\
$^{22}$Also at Istinye University, Istanbul, Turkey\\
$^{23}$Also at an international laboratory covered by a cooperation agreement with CERN\\
$^{24}$Also at The University of the State of Amazonas, Manaus, Brazil\\
$^{25}$Also at University of Hamburg, Hamburg, Germany\\
$^{26}$Also at RWTH Aachen University, III. Physikalisches Institut A, Aachen, Germany\\
$^{27}$Also at Bergische University Wuppertal (BUW), Wuppertal, Germany\\
$^{28}$Also at Brandenburg University of Technology, Cottbus, Germany\\
$^{29}$Also at Forschungszentrum J\"{u}lich, Juelich, Germany\\
$^{30}$Also at CERN, European Organization for Nuclear Research, Geneva, Switzerland\\
$^{31}$Also at HUN-REN ATOMKI - Institute of Nuclear Research, Debrecen, Hungary\\
$^{32}$Now at Universitatea Babes-Bolyai - Facultatea de Fizica, Cluj-Napoca, Romania\\
$^{33}$Also at MTA-ELTE Lend\"{u}let CMS Particle and Nuclear Physics Group, E\"{o}tv\"{o}s Lor\'{a}nd University, Budapest, Hungary\\
$^{34}$Also at HUN-REN Wigner Research Centre for Physics, Budapest, Hungary\\
$^{35}$Also at Physics Department, Faculty of Science, Assiut University, Assiut, Egypt\\
$^{36}$Also at Punjab Agricultural University, Ludhiana, India\\
$^{37}$Also at University of Visva-Bharati, Santiniketan, India\\
$^{38}$Also at Indian Institute of Science (IISc), Bangalore, India\\
$^{39}$Also at Amity University Uttar Pradesh, Noida, India\\
$^{40}$Also at UPES - University of Petroleum and Energy Studies, Dehradun, India\\
$^{41}$Also at IIT Bhubaneswar, Bhubaneswar, India\\
$^{42}$Also at Institute of Physics, Bhubaneswar, India\\
$^{43}$Also at University of Hyderabad, Hyderabad, India\\
$^{44}$Also at Deutsches Elektronen-Synchrotron, Hamburg, Germany\\
$^{45}$Also at Isfahan University of Technology, Isfahan, Iran\\
$^{46}$Also at Sharif University of Technology, Tehran, Iran\\
$^{47}$Also at Department of Physics, University of Science and Technology of Mazandaran, Behshahr, Iran\\
$^{48}$Also at Department of Physics, Faculty of Science, Arak University, ARAK, Iran\\
$^{49}$Also at Italian National Agency for New Technologies, Energy and Sustainable Economic Development, Bologna, Italy\\
$^{50}$Also at Centro Siciliano di Fisica Nucleare e di Struttura Della Materia, Catania, Italy\\
$^{51}$Also at Universit\`{a} degli Studi Guglielmo Marconi, Roma, Italy\\
$^{52}$Also at Scuola Superiore Meridionale, Universit\`{a} di Napoli 'Federico II', Napoli, Italy\\
$^{53}$Also at Fermi National Accelerator Laboratory, Batavia, Illinois, USA\\
$^{54}$Also at Lulea University of Technology, Lulea, Sweden\\
$^{55}$Also at Consiglio Nazionale delle Ricerche - Istituto Officina dei Materiali, Perugia, Italy\\
$^{56}$Also at Institut de Physique des 2 Infinis de Lyon (IP2I ), Villeurbanne, France\\
$^{57}$Also at Department of Applied Physics, Faculty of Science and Technology, Universiti Kebangsaan Malaysia, Bangi, Malaysia\\
$^{58}$Also at Consejo Nacional de Ciencia y Tecnolog\'{i}a, Mexico City, Mexico\\
$^{59}$Also at Trincomalee Campus, Eastern University, Sri Lanka, Nilaveli, Sri Lanka\\
$^{60}$Also at Saegis Campus, Nugegoda, Sri Lanka\\
$^{61}$Also at National and Kapodistrian University of Athens, Athens, Greece\\
$^{62}$Also at Ecole Polytechnique F\'{e}d\'{e}rale Lausanne, Lausanne, Switzerland\\
$^{63}$Also at Universit\"{a}t Z\"{u}rich, Zurich, Switzerland\\
$^{64}$Also at Stefan Meyer Institute for Subatomic Physics, Vienna, Austria\\
$^{65}$Also at Laboratoire d'Annecy-le-Vieux de Physique des Particules, IN2P3-CNRS, Annecy-le-Vieux, France\\
$^{66}$Also at Near East University, Research Center of Experimental Health Science, Mersin, Turkey\\
$^{67}$Also at Konya Technical University, Konya, Turkey\\
$^{68}$Also at Izmir Bakircay University, Izmir, Turkey\\
$^{69}$Also at Adiyaman University, Adiyaman, Turkey\\
$^{70}$Also at Bozok Universitetesi Rekt\"{o}rl\"{u}g\"{u}, Yozgat, Turkey\\
$^{71}$Also at Marmara University, Istanbul, Turkey\\
$^{72}$Also at Milli Savunma University, Istanbul, Turkey\\
$^{73}$Also at Kafkas University, Kars, Turkey\\
$^{74}$Now at Istanbul Okan University, Istanbul, Turkey\\
$^{75}$Also at Hacettepe University, Ankara, Turkey\\
$^{76}$Also at Erzincan Binali Yildirim University, Erzincan, Turkey\\
$^{77}$Also at Istanbul University -  Cerrahpasa, Faculty of Engineering, Istanbul, Turkey\\
$^{78}$Also at Yildiz Technical University, Istanbul, Turkey\\
$^{79}$Also at School of Physics and Astronomy, University of Southampton, Southampton, United Kingdom\\
$^{80}$Also at IPPP Durham University, Durham, United Kingdom\\
$^{81}$Also at Monash University, Faculty of Science, Clayton, Australia\\
$^{82}$Also at Universit\`{a} di Torino, Torino, Italy\\
$^{83}$Also at Bethel University, St. Paul, Minnesota, USA\\
$^{84}$Also at Karamano\u {g}lu Mehmetbey University, Karaman, Turkey\\
$^{85}$Also at California Institute of Technology, Pasadena, California, USA\\
$^{86}$Also at United States Naval Academy, Annapolis, Maryland, USA\\
$^{87}$Also at Ain Shams University, Cairo, Egypt\\
$^{88}$Also at Bingol University, Bingol, Turkey\\
$^{89}$Also at Georgian Technical University, Tbilisi, Georgia\\
$^{90}$Also at Sinop University, Sinop, Turkey\\
$^{91}$Also at Erciyes University, Kayseri, Turkey\\
$^{92}$Also at Horia Hulubei National Institute of Physics and Nuclear Engineering (IFIN-HH), Bucharest, Romania\\
$^{93}$Now at another institute formerly covered by a cooperation agreement with CERN\\
$^{94}$Also at Texas A\&M University at Qatar, Doha, Qatar\\
$^{95}$Also at Kyungpook National University, Daegu, Korea\\
$^{96}$Also at Yerevan Physics Institute, Yerevan, Armenia\\
$^{97}$Also at another international laboratory covered by a cooperation agreement with CERN\\
$^{98}$Also at Imperial College, London, United Kingdom\\
$^{99}$Also at Institute of Nuclear Physics of the Uzbekistan Academy of Sciences, Tashkent, Uzbekistan\\
$^{100}$Also at another institute formerly covered by a cooperation agreement with CERN\\
\end{sloppypar}
\end{document}